\documentclass[figures]{epl}

\title{Jamming transitions in a schematic model of suspension rheology}
\shorttitle{Jamming transitions}
\author{C. B. Holmes \inst{1} \and M. Fuchs \inst{2} \and M. E. Cates \inst{1}}
\shortauthor{C. B. Holmes \etal}
\institute{
  \inst{1} School of Physics, The University of Edinburgh  - JCMB, The Kings Buildings, Edinburgh, EH9 3JZ, Scotland.\\
  \inst{2} Institut Charles Sadron - 6 rue Boussingault, 67083 Strasbourg Cedex, France.}

\begin{document}

\maketitle

\begin{abstract}
We study the steady-state response to applied stress in a simple scalar model of sheared colloids. Our model is based on a schematic (F2) model of the glass transition, with a memory term that depends on both stress and shear rate. For suitable parameters, we find transitions from a fluid to a nonergodic, jammed state, showing zero flow rate in an interval of applied stress. Although the jammed state is a glass, we predict that jamming transitions have an analytical structure distinct from that of the conventional mode coupling glass transition.
The static jamming transition we discuss is also distinct from
hydrodynamic shear thickening.

\end{abstract}
When subjected to an applied shear stress, concentrated hard-sphere
colloids are known to shear-thicken \cite{Frith,Laun,Bender,
O'Brien+Mackay}. An
understanding of this in terms
of hydrodynamic interactions has been developed \cite{Ball+Melrose95, BradyReview}. Recently, a different
(although seemingly related) phenomenon has been found. Bertrand
and co-workers \cite{Bertrand2002} report an experiment in which
shearing a concentrated suspension induces a transition from a fluid to a
metastable solid {\it which persists after cessation of
shear}. Vibration of the sample restores it to
a fluid state. 

Because the solid paste persists in the absence of flow, such `static
jamming' cannot be explained by hydrodynamic forces alone. An
alternative view is that after cessation of flow the material remains
stressed; this stress is what maintains the arrest. Indeed, jammed
suspensions often take on a lumpy dry appearance as particles protrude
partially from the fluid surface (a dilatancy effect). This entails
large capillary stresses which might maintain the jam; lumps of paste
removed from a rheometer would then remain solid. Vibration could
destroy the jammed structure and restore both the fluidity and a wet
appearance, as is observed \cite{Bertrand2002,warrenpc, hawpc}. Rather than model all this directly, we focus here on the simpler case of stress-induced jamming within an idealised rheometer, with no free surfaces.

It was recently proposed that jamming in granular
materials \cite{forcechains} might be closely connected to the glass transition \cite{LiuNagel,
Liu+Nagel-Book}. If so, the jammed colloidal state
found by Bertrand {\it et al} could resemble a glass, albeit one which is
stressed and hence anisotropic. (Colloidal glasses are well-documented \cite{PuseyVanMegan1996}.) A speculative scenario for
the formation of a shear-induced glass is as follows: the applied shear stress alters the structure of the material, initially through shear flow and hydrodynamics. But as the material thickens, its structure becomes more susceptible to dynamical arrest, because an increased number of close interparticle contacts hinders diffusion. This promotes a nonergodicity transition, at concentrations somewhat lower than would be required without the stress. In the arrested state, flow ceases but stress remains; the stress is now sustained by interparticle and/or entropic forces rather than hydrodynamics.
 
The consideration of such a phenomenon is complicated by the tensorial nature of applied stresses; a system jammed by one component of an applied stress might be refluidised by another \cite{forcechains}. In the
current work, we ignore these tensorial complications, and treat jamming as a
generalised glass transition within a scalar model. (For a somewhat complementary approach, see \cite{Head2001}.) This model builds
on the ideas of the mode-coupling theory (MCT) of the glass
transition, but introduces two new features: stress-induced arrest,
and strain-induced memory loss. The combination of these features
allows new shear-thickening and jamming scenarios to emerge. Fuller
details of our calculations will appear elsewhere \cite{Holmes-long}.  

For a comprehensive description of MCT, see \cite{Goetze}. We outline only those aspects pertinent to the present work.
The central quantities of the theory are density fluctuations at wavevector ${\mathbf q}$,   $\,\delta\rho({\mathbf q},t)$. The correlators of these quantities, $\phi_{\mathbf q}(t)\equiv\langle\delta\rho({\mathbf q},t)\delta\rho(-{\mathbf q},0)\rangle/\langle|\delta\rho({\mathbf q})|^2\rangle$ may be measured in scattering experiments, and provide a description of the system's dynamics. In a liquid, the system is ergodic and  $\phi_{\mathbf q}(t)$ decays to zero with time for all ${\bf q}$. In a glass it does not: $\lim_{t\to\infty} \phi_{\bf q}(t) =f_{\bf q}>0$, where $f_{\bf q}$ is a nonergodicity parameter characteristic of the arrested structure.  A finite $f_{\bf q}$ represents  the inability of the structure to relax on lengthscale $\sim 2\pi/q$, preventing an initial fluctuation from fully decaying. 

Equations of motion can be found for the correlators $\phi_{\mathbf q}(t)$. On making approximations suitable for a colloidal system, and dropping $\mathbf{q}$-subscripts, the result is \cite{GoetzeEssentials}:
\begin{equation}
\phi(t)+\tau_o\dot{\phi}(t)+\int_0^tm(t-t^\prime)\dot{\phi}(t^\prime)\,dt^\prime=0,\label{EOM}
\end{equation}
where $\tau_o$ sets the timescale for the microscopic dynamics. Here $m(t-t^\prime)$ is the {\it memory function}, and describes a retarded friction effect which, in the colloidal glass transition, arises by caging of a particle by its neighbours. In MCT, the memory function is found approximately by integrating (over wavevectors) a quadratic product of correlators, with coupling constants that depend on the static structure factor $S(q) = (1/N)\langle|\delta\rho({\mathbf q})|^2\rangle$ of the system. This  
approximation means that anharmonic interactions between density fluctuations are ignored.

Generalisation of the wavevector-dependent MCT to sheared systems  is
challenging, and so far has shown only shear-thinning behavior
\cite{Fuchs+Cates2002, Miyazaki, Indrani}.
Analogous behaviour is seen in mean field spin models of driven glasses \cite{Berthier}. To address the more complicated issue of jamming, a simpler starting point is needed.  A promising one lies with `schematic models' which capture many key features of the MCT glass transition. These schematic models consider a single correlator, rather than the infinite set $\{\phi_{\bf q}(t)\}$. The memory function is then written as a polynomial of the correlator, with coefficients (coupling constants) that schematically represent system variables. Larger coupling constants correspond to higher densities or lower temperatures.
We choose to build upon a simple schematic model (the `F2 model') for which $m(t)=v\phi^2(t)$. This model has an idealised glass transition at $v=4$ \cite{Leutheusser}: for $v<4$, the only admissable solutions have $f=0$, while for larger $v$ nonergodic solutions, with $f>0$, appear.

To incorporate shearing into this picture we need to account for two
effects. Firstly, flow suppresses memory: particles separated by a
small distance in the flow gradient direction become well separated
for times beyond $1/\dot\gamma$ (with $\dot\gamma$ the shear
rate). Beyond this timescale, any system at nonzero flow rate should lose memory of previous configurations, so that ergodicity
is restored \cite{Sollich98,Berthier}. (In ${\mathbf q}$-space, the
important fluctuations, which have wavevectors near the peak of
$S(q)$, are advected on this timescale to higher $q$ where they decay
rapidly. Hence cages are destroyed, and even those fluctuations that
are not directly advected, become ergodic \cite{Fuchs+Cates2002}.)
The second effect is jamming: the fact that stress (as distinct from
flow) can {\em promote} arrest. Stress can hinder diffusion by
distorting cages and creating more close contacts between particles,
particularly if these tend to arrange into a load-bearing structure
\cite{forcechains}. (This idea does not exclude a strong transient role for hydrodynamic effects in determining what structures actually do arise.) Such physics is
unlikely to be captured within the harmonic approximation normally
used in full MCT. To capture it, one might make a harmonic expansion
using the $S(q)$ of some anisotropic reference state in which the
thermodynamic stress was nonzero: at the schematic level, we should
thus allow a stress dependence of the coefficients, alongside an
explicit memory loss arising from nonzero $\dot \gamma$. 

To allow for both these effects, we model the memory function under shear as follows:
\begin{equation}
m(t)= (v_o+\alpha\sigma)\exp(-\dot\gamma t)\phi^2(t).
\label{first}
\end{equation}
Here the parameter $v_o$ represents the system's tendency to arrest in
the absence of any external forcing, and $\alpha$ represents the degree to which
this intrinsic memory is enhanced by a shear stress of magnitude $\sigma$. 
(Clearly the interesting case for jamming systems is $\alpha > 0$.)
At
small $\sigma$ one could argue for a quadratic dependence (since
positive and negative shear stresses should be equivalent) but we
shall find that the interesting behavior of the model in fact lies in
a window of stress where $\sigma$ is not small. (The chosen $\sigma$
dependence can then be interpreted as a linearisation about some point
near the middle of this window.)  
The exponential form chosen for flow-induced memory loss is somewhat
arbitary, but the results do not depend much upon this
choice: the important part of this function involves  $\dot\gamma t
\ge 1$ (so that the apparent nonanalytic dependence at small flow
rates is again immaterial). Another form, $m(t)\propto
1/(1+(\dot\gamma t)^2)$, which is better motivated by microscopic
considerations \cite{Fuchs+Cates2002}, yields qualitatively similar
numerical results, but the exponential is preferable in analytic
work \cite{Holmes-long}.

In steady flow the stress $\sigma$ and strain rate $\dot\gamma$ are not independent, but are related through a viscosity $\eta(\dot\gamma)\equiv \sigma/\dot\gamma$ which depends on the state of the material. To close our schematic model, we need a prescription for this viscosity. In linear response, the viscosity may be expressed as the time integral of a stress correlator \cite{HansenMcDonald}. In our simplified model, we have only one correlator, and so write 
\begin{equation}
\eta=\int_0^\infty \phi(t)\,\mathrm{dt}=\tau,
\label{second}
\end{equation}
where we choose units so that $\eta$ equates to a characteristic relaxation time $\tau$.
Equation \ref{second} is again somewhat ad-hoc, since we are not studying
the linear response regime; but equating $\eta$ to an integral of
$\phi(t)^2$ (say) would give very similar results. More generally,
according to the full MCT \cite{Goetze} the viscosity
diverges like the relaxation time of a typical correlator as one
approaches a colloidal glass transition (a correspondence that is not
lost under shear \cite{Fuchs+Cates2002}); Eq.\ref{second} captures
this. It means that dynamical arrest (nonzero $f$) implies a divergent
viscosity and hence zero shear rate. This is fully consistent with
arguments made above that finite $\dot\gamma$ prohibits arrest. (Note
that the shear rate refers to the steady state value, so this
does not exclude the possibility of sublinear creep of the strain, with vanishing shear rate at late times.)

Our model is completely defined by Eqs. \ref{EOM}, \ref{first} and
\ref{second}. It has been solved numerically by adapting an
established algorithm for MCT equations \cite{Matthias1991}. This
allows iteration (from above or below) of the relaxation time $\tau$
(or equivalently $\dot\gamma$) to a self-consistent solution for fixed
$v_o$, $\alpha$ and $\sigma$. Below, when we refer to stability of the
solutions, we mean stability under this iteration. (Physical stability
is discussed at the end.) The numerical results were also checked by
analysis of limiting cases \cite{Holmes-long}. 
Note that unstable solutions (in the sense just defined) also exist:
by inspection, for $\dot\gamma=0$, the model reduces to the standard
F2 model, with coupling $v=v_o+\alpha\sigma$. Therefore, for nonzero
$\alpha$, a nonergodic solution at zero shear rate always formally
exists for $\sigma$ sufficiently large ($v_o+\alpha\sigma >
4$). Numerically however, it was found that this state is unstable
with respect to an ergodic solution whenever the latter
exists. However, for some parameter choices, there is no ergodic
solution within a certain window of stress. Here the nonergodic jammed
state, with finite stress but zero flow rate, is stable. 

\begin{figure}
\onefigure[scale=0.4]{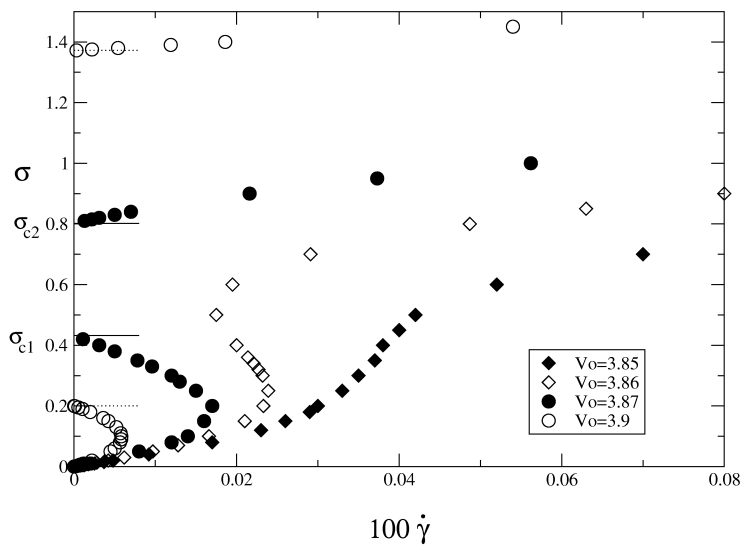}
\caption{Flow curves, stress $\sigma$ versus shear rate $\dot\gamma$, for $\alpha=0.95$. For the two largest values of $v_o$,
it appears that for a window in $\sigma$, the relaxation time has
diverged. Analytic calculations of the limits of this window are
indicated as horizontal lines near the stress axis. These values of
the stress are dubbed $\sigma_{c1}$ and $\sigma_{c2}$, as
shown here for one of the parameter sets. Larger shear rates (which have been made
dimensionless by mutiplication with the `bare' relaxation time $\tau_o$) are
found at larger stresses. 
}\label{flowcurves}
\end{figure}

The resulting `full jamming'  scenario (see also \cite{Head2001})
forms part of a wider range of rheological
behavior. Fig. \ref{flowcurves} shows three thickening
scenarios, dependent on model parameters; $v_o$ is chosen close to the
quiescent glass transition and $\alpha$ is chosen close to unity. Upon
increasing $v_o$ there is a progression from a monotonic, continuously
shear-thickening curve, via a nonmonotonic S-shaped curve, to a curve
that extends right back to the vertical axis. It is known that any
flow curve with negative slope is unstable to shear banding
\cite{Olmsted}; hence for the latter two scenarios, in any experiment
at controlled shear rate, the stress would be expected to jump
discontinuously from the lower to the upper branch before reaching the
point of infinite slope. Such discontinuous shear thickening is widely
reported \cite{Frith,Laun,Bender,O'Brien+Mackay} and usually
attributed to hydrodynamic interactions \cite{BradyReview}. Our work
suggests that, at least in some systems, this may not be the only
mechanism at work. In particular, Fig. \ref{flowcurves} shows shear
thickening at Peclet numbers $\dot\gamma\tau_0 \sim 10^{-4}$, rather
than values of order ten predicted by hydrodynamics. 
We will discuss this in greater depth in a future paper \cite{Holmes-long}.

\begin{figure}
\onefigure[scale=0.4]{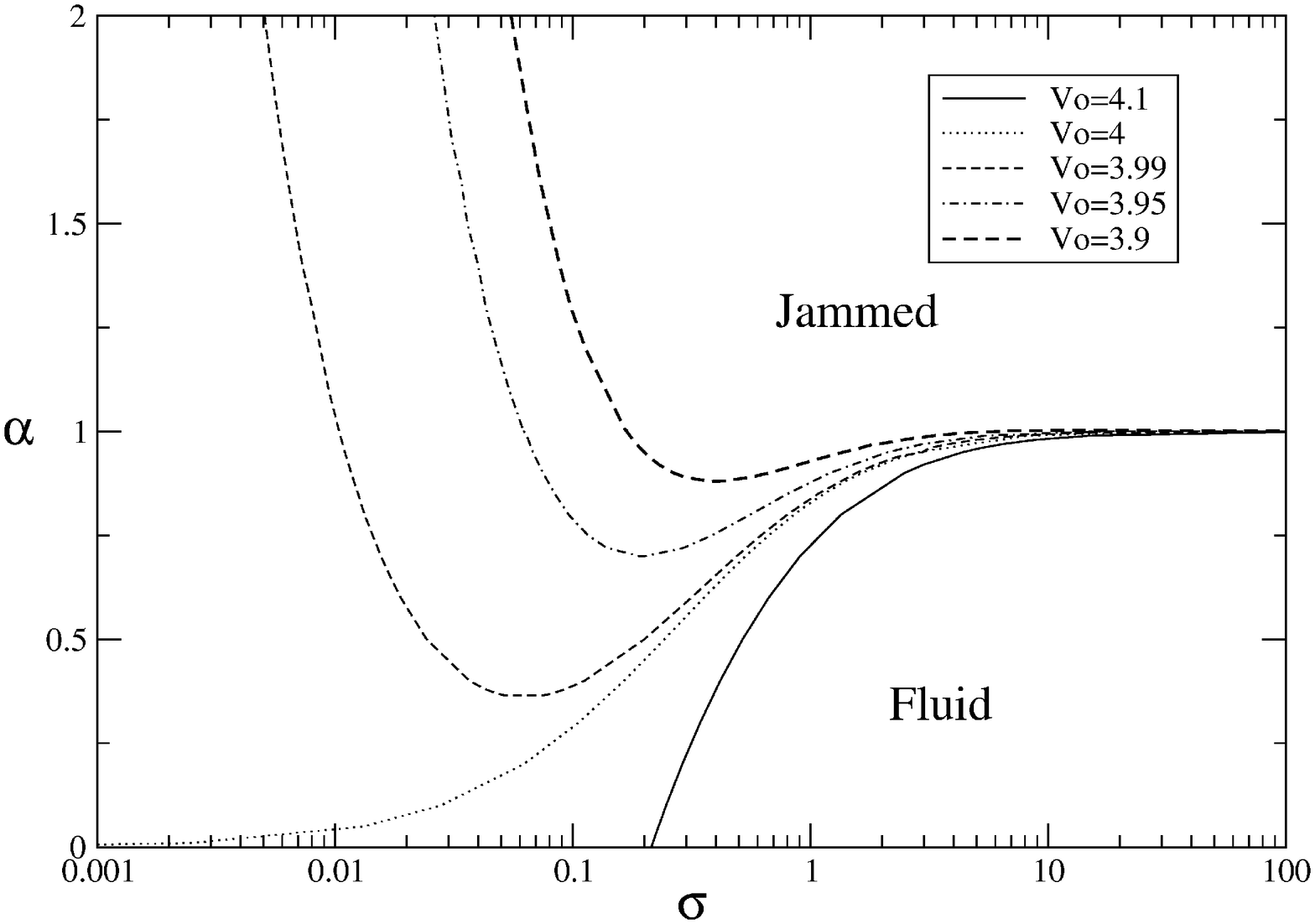}
\caption{`Phase diagrams' for the model for various $v_o$. The lines denote transitions in the $(\alpha,\sigma)$-plane. All states below the curve for a given value of $v_o$ are fluid states, whilst those above (and on) the line are nonergodic, jammed states.}\label{pdiagram}
\end{figure}

For the largest values of the parameter $v_o$, in Fig. \ref{flowcurves},
there is a range of stress for which the shear rate returns to zero:
there is then no ergodic solution, and the jammed state is
stable. This represents full, static jamming. 
The lower and upper endpoints $\sigma_{c1}$ and $\sigma_{c2}$ of the
stable jammed state represent distinct jamming transitions. Their
critical stresses obey
\begin{equation}
f_c\left[\left(v_o+\alpha\sigma_c\right)f_c-2\right]=\sigma_c,\label{transitions}
\end{equation}
where $f_c$ is given by the largest solution of $\frac{f_c}{1-f_c}=(v_o+\alpha\sigma_c)f_c^2$ .
Such transitions exist provided that both $v_o$ and $\alpha$ are
sufficiently large. The meaning of $v_o$ is clear:
it represents the tendency of an unsheared suspension to arrest, and
is controlled by the colloid concentration and interactions \cite{Goetze}. The
experimental meaning of $\alpha$ is less clear. Bertrand {\it
et al} \cite{Bertrand2002} found that, for concentrations
below a certain value, their samples showed ordinary thickening, whilst above this
value the shear-induced solid was seen. The behaviour illustrated in
Fig. \ref{flowcurves} is reminiscent of this. Note that the
re-fluidisation under increasing stress is dependent upon the value of
$\alpha$: if this parameter is sufficiently large (for a given value
of $v_o$) this re-fluidisation is not present. This is illustrated in
a `phase diagram' of the model, dividing parameter space into ergodic
and nonergodic regions (Fig. \ref{pdiagram}). This shows that at large enough stresses, jammed (fluid) states arise for $\alpha > 1$ ($\alpha <1$). However, for particle densities close to but below the quiescent glass transition, for $\alpha<1$ the system jams in an intermediate window of stress.

We now examine the generic behavior at the jamming transitions within
our model, and compare this to the glass transition as conventionally
addressed in MCT \cite{Goetze}. Note that, because of
an expected shear-banding instability associated with the decreasing part of the
flow curve \cite{Olmsted}, the
lower jamming transition at $\sigma_{c1}$ might be more complicated to
access experimentally than the upper one at $\sigma_{c2}$. The latter
should always be accessible by raising the stress in the fluid until
discontinuous shear thickening occurs (taking the system to
the upper branch of the flow curve) and then reducing the stress
through the transition at $\sigma_{c2}$. This practical difference
aside, we find similar behavior at both transitions for all the
features discussed below.

\begin{figure}
\onefigure[scale=0.6]{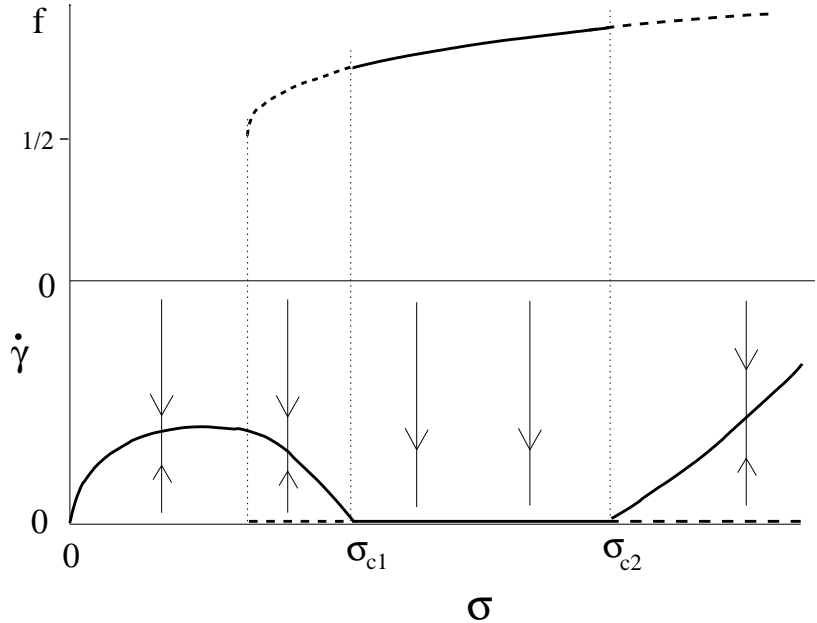}
\caption{Schematic variation of $\dot\gamma$ and $f$ on varying
$\sigma$. Solid (dashed) lines denote stable (unstable)
solutions. Arrows indicate flow under iteration. Note that the
ergodic, flowing solution remains stable beyond the bifurcation point
where a nonergodic solution first arises: {\it ie},
$v_o+\alpha\sigma_{c1}>4$. Hence there are no nonanalyticities after
the discontinuity in $f$ at the lower or upper jamming
transition. }\label{schematic figure}
\end{figure}

The standard MCT glass transition is associated with a bifurcation of
$f$ in parameter space: at sub-critical coupling, there is no solution
with a nonzero  $f$. At the transition point, there is a bifurcation
in the solution adopted by the dynamics for $t\to\infty$. This leads to a jump in $f$, from zero to
a finite value, followed by a {\em nonanalytic} variation as the
coupling is increased. At our jamming transitions, there is no
bifurcation in $f$ at the transition point, and so the jump in $f$ is
followed by a linear variation with stress. Indeed, the jamming
transitions in our model mark the disappearance of the ergodic
(flowing) solution -- stable whenever it exists -- rather than the
first appearance of the nonergodic (arrested) one. The latter is where
the bifurcation occurs, and while in the MCT of conventional glasses
this coincides with disappearance of the ergodic solution, here it
does not. The bifurcation and stability behavior of our model is
summarised in Figure \ref{schematic figure}.
There is a second, closely related, difference between the jamming
transitions described here and the conventional MCT glass
transition. In the vicinity of a conventional transition, there are
two divergent timescales, those of the (short time) $\beta$ and (long
time) $\alpha$ relaxation processes. In contrast, at the jamming
transitions described here, there is only one. Again, this is because
the transition occurs at a different point in parameter space to the
underlying glass transition.

Throughout the above discussion, we have tacitly assumed that the
stability of our solutions under iteration also governs the physical
selection mechanism: a flowing state is preferred to a jammed one,
whenever both exist. Physically this is not obvious, but plausible;
although the iteration does not map directly onto a dynamical
evolution, it does suggest that any transient violation of
Eq.\ref{second}, leading temporarily to an infinitesimal shear rate in
the nonergodic state, will carry the system towards a steadily flowing
solution unless $\sigma_{c1} < \sigma < \sigma_{c2}$. Nonetheless,
recall that a formal nonergodic solution does exist at zero shear rate
whenever $v_o+\alpha\sigma >4$. Were this to be physically stable and
the flowing solution unstable, we would have only one jamming
transition, showing all the behaviour of a conventional glass
transition. But reports of flowing colloids at high stresses in
discontinuously shear-thickening systems \cite{Laun,Frith,Bender,
O'Brien+Mackay} make that outcome unlikely. A more plausible (though
equally speculative) scenario is hysteresis: it might be posssible to
maintain a metastable jammed state outside the range
$\sigma_{c1}<\sigma<\sigma_{c2}$ by starting within that range and
carefully changing the stress. The model predicts no limit to this at
high stresses, but at low ones, metastability cannot extend below
$\sigma_{c0}\equiv (4-v_o)/\alpha$.

In this Letter we have presented a simple model of jamming in
suspensions, based on a schematic mode-coupling model of the glass
transition. Our model exhibits transitions to a nonergodic state,
which has properties similar to that of the idealised glass state of
MCT \cite{Goetze}, but (assuming that the iterative stability of
solutions also controls the physical behavior) the jamming transitions
that we find have a novel structure. They are not controlled directly
by a bifurcation of the nonergodicity parameter, but instead by the
disappearance of an ergodic, fluid phase within a window of stress.
Aspects of the model appear consistent with recent experiments
\cite{Bertrand2002}, although more work is required to clarify the nature of static jamming in dense colloids under stress. We hope that our work will stimulate such experiments.

\acknowledgments
We thank Patrick Warren, Mark Haw, J\'er\^ome Bibette, Norman Wagner
and Thomas Voigtmann for useful discussions. MF was supported by the
DFG, grant Fu 309/3. CH thanks Unilever and EPSRC for a CASE
studentship.

\end{document}